# Design of an achromatic zoom metalens doublet in the visible

TIE HU, [1,†] XING FENG, [1,†] YUNXUAN WEI, [2] SHENGQI WANG, [1] YUHONG WEI, [1] ZHENYU YANG, [1*] AND MING ZHAO[1*]

[1]Nanophotonics Laboratory, School of Optical and Electronic Information, Huazhong University of Science and Technology, Wuhan 430074, China
[2]Ming Hsieh Department of Electrical and Computer Engineering, University of Southern California, Los Angeles, California 90089, USA
*Corresponding author: zhaoming@hust.edu.cn , zyang@hust.edu.cn

**Zoom metalens doublets, featuring ultra-compactness, strong zoom capability and CMOS compatibility, exhibit unprecedented advantages over the traditional refractive zoom lens. However, the huge chromatic aberration narrows the working bandwidth, which limits their potential applications in broadband systems. Here, by globally optimizing the phase profiles in the visible, we designed and numerically demonstrated a Moire lens based zoom metalens doublet that can achromatically work in the band of 440-640 nm. Such a doublet can achieve a continuous zoom range from 1× to 10×, while also maintaining high focusing efficiency up to 86.5% and polarization insensitivity. © 2022 Optical Society of America**

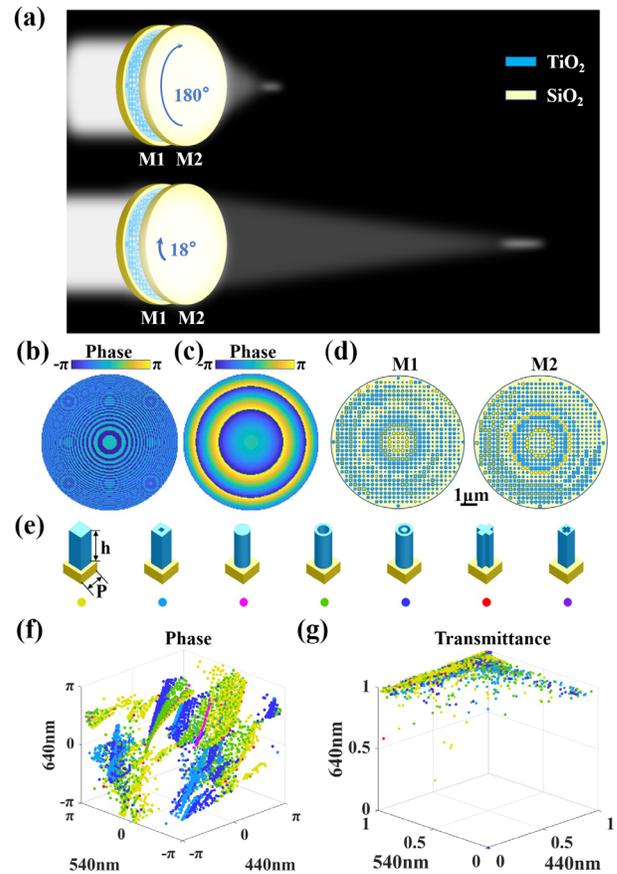

**Fig. 1.** Principle of the achromatic zoom metalens doublet and design of the meta-atom. (a) Schematic of the doublet. Arbitrary zoom power from ±1 to ± ∞ can be realized by adjusting the rotation angle of the composed metasurfaces. Yellow and blue colors represent $SiO_2$ and $TiO_2$, respectively. (b-c) The quantized phase profiles of the doublet under the rotation angles of π and π/10. (d) Top view of the central areas of the two designed metasurfaces. The scale bar is 1 μm. (e) Illustration of the seven types of meta-atoms. The (f) phase shift, and (g) transmittance of the meta-atoms. Each color corresponds to the meta-atom marked by the point with the same color in (e).

**Introduction.** Metasurfaces, well-known for their flexible manipulation of phase, amplitude, and polarization at the subwavelength scale, have been widely studied and applied from biosensors [1], polarization detectors [2, 3] to zoom lenses [4, 5]. Owning to their planar geometries, zoom metalenses possess extreme compactness and integration which are impossible for traditional lens groups. To obtain the zoom ability, researchers have put intensive effort into making metalens tunable, for which mechanic tuning is one of the most effective approaches [6-11]. For instance, Shane et al. and Wei et al. respectively realized mechanically tunable metalens doublets based on the Alvarez lens [8] and the Moiré lens principles [10], and separately demonstrated the zoom ability of 3× and 18×. Notably, the zoom power of the latter can be changed from ±1 to theoretically ± ∞ through relatively rotating the two metasurfaces, therefore keeping the optical system compact and miniature. Moreover, compared to designs of the Alvarez lens, the Moiré lens can change the focal length with a fixed effective lens aperture, thus leading to higher efficiency. However, the strong chromatic aberration, arising from the wavelength-dependent response of the meta-atom and the diffraction, prevents these designs from practical applications which always require uniform focusing for broad bands.

The chromatism of a metalens can be fully compensated by carefully controlling the group delay and the group delay dispersion of every meta-atom [12, 13]. However, the insufficient dispersion of the subwavelength structure introduces a severe trade-off between the numerical aperture, the diameter and the working bandwidth [14]. Alternatively, the focal length deviations at discrete wavelengths are supposed to be suppressed by multiwavelength metasurfaces enabled by spatial multiplexing [15, 16], and multiple resonances [17, 18].

Here, an achromatic zoom metalens doublet is realized by globally optimizing the phase of the Moiré lens at discrete wavelengths. Our previous work has demonstrated the concept of rotational zoom doublet at the wavelength of 1550 nm [10], the present paper is original in verifying an achromatic zoom metalens doublet in the visible. By changing the rotation angle between the two metasurfaces, we numerically verify a polarization-insensitive metalens doublet with a zoom capacity up to 10×. To characterize the focusing performance, we further study the average focal length error and focusing efficiency. The maximum average focal length error is below 6.28% and the efficiency is up to 86.2%.

**Design.** Fig. 1(a) illustrates the principle of the achromatic zoom metalens doublet. Ideally, by continuously adjusting the rotation angle between the two metasurfaces (denoted as M1 and M2), this doublet can realize achromatic and tunable focusing in the visible. For example, the zoom powers of 1× and 10× are respectively realized by adjusting the rotation angles of 180° and 18°. Following the Moiré lens principle, the phase distributions of the two metasurfaces in the doublet should depend on the rotation angle. Rather than selecting a hyperbolic phase profile, we choose the phase profile of a spherical lens due to smaller phase errors [10]. The phase profiles of M1 and M2 are described as

$$\varphi_1(r,\theta_0,\lambda) = \text{round}(\pi r^2/\lambda F_0)\,\theta_0 + C(r,\lambda), \quad (1)$$

$$\varphi_2(r,\theta_0,\lambda) = -\varphi_1(r,\theta_0,\lambda) = -\text{round}(\pi r^2/\lambda F_0)\theta_0 - C(r,\lambda), \quad (2)$$

where $r$ and $\theta_0$ are the radial coordinates, $\lambda$ is the working wavelength, and $F_0$ is a reference focal length. The operator round(…) replaces its parameter with the nearest integer. Hence, this quantization operator ensures the phase change with an integer multiple of $2\pi$ when $\theta_0$ varies from 0 to $2\pi$. Otherwise, the discontinuity around $\theta_0 = 2\pi$ leads to unwanted phase errors. $C(r,\lambda)$ is chosen as a function of $r$ and $\lambda$ to serve as an optimization term and correct distortions due to the finite distance between M1 and M2. Combining the two metasurfaces, the phase profile of the doublet is

$$\varphi_{\text{lens}}(r,\theta,\lambda) = \varphi_1(r,\theta_0,\lambda) + \varphi_2(r,\theta_0+\theta,\lambda) = -\text{round}(\frac{\pi}{\lambda F_0}r^2)\theta. \quad (3)$$

Comparing with the phase distribution of an ideal spherical lens (given by $\varphi_S = -\pi r^2/\lambda F$), $\varphi_{\text{lens}}$ depicted in Figs. 1(b) and 1(c) are the quantized results of $\varphi_S$. We can find that the focal length of the doublet can be expressed as

$$F(\theta) = F_0/\theta. \quad (4)$$

Here, the zoom power of the doublet is defined as $K=\pi/\theta$. The mentioned quantization causes lower efficiency. Based on the theory of multilevel diffractive lens [19], the efficiency is $\eta = (\text{sinc}(1/N))^2$, where $N=2K$ is the quantized level. The ideal efficiency is positively correlated with zoom power.

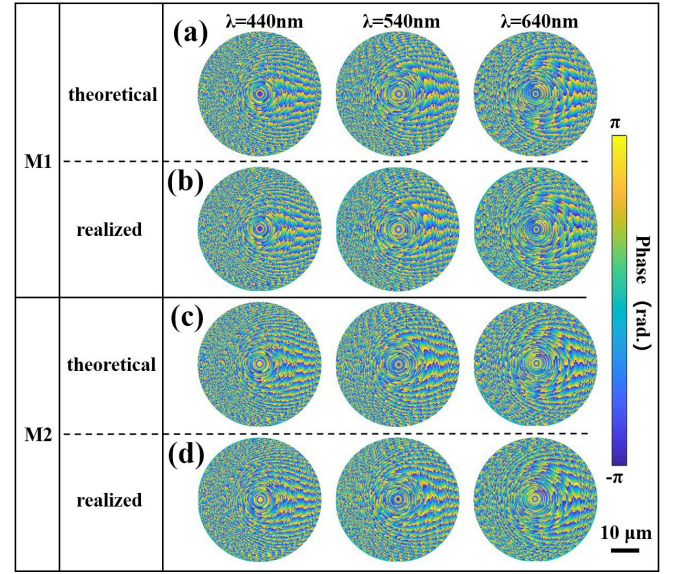

**Fig. 2.** The phase profiles of the two metasurfaces in the doublet. The theoretical (a) and realized (b) phase profiles of M1. The theoretical (c) and realized (d) phase profiles of M2. Figures from the left channel to the right channel correspond to the designed wavelengths. The scalar bar is 10 μm.

To achieve the above phase profiles, the metasurfaces are composed of $TiO_2$ or air-hole meta-atoms, as depicted in Fig. 1 (d). Fig. 1 (e) shows that seven types of meta-atoms are used to construct the phase library $\Phi(\lambda)$, such as square, circle, and ring, which are marked by different colored dots. The meta-atom is composed of a four-fold symmetrical $TiO_2$ or air-hole nanopillar based on the square $SiO_2$ substrate, with a period of 300 nm and a height of 900 nm. This symmetrical geometry of meta-atoms can ensure polarization insensitivity of the doublet. Here, the finite time domain difference (FDTD) method is used to calculate the optical response of the meta-atoms under the x linear polarization light incidence from the substrate. Figs. 1 (f) and 1 (g) respectively show that $\Phi(\lambda)$ covers from $-\pi$ to $\pi$, and the transmittance mainly approaches unity at the three design wavelengths. Then, we only optimize the phase profiles of the two metasurfaces without considering their transmittances.

For simplicity, $C(r,\lambda)$ is set as a rotationally symmetric function with $r$ during the optimization. Only $C(r,\lambda)$ along the radial direction needs to be optimized at every designed wavelength. The core of the optimization method is to reduce the average phase errors between the ideal phase profiles and phase library $\Phi(\lambda)$ at the designed wavelengths. Here, the figure of metric is defined as

$$FOM_i = \min\left\{\left\{\sum_{\theta_0=0}^{2\pi}\sum_{r=0}^{R}\min\left[\sum_{\lambda=\lambda_{\min}}^{\lambda_{\max}}|\varphi_1(r,\lambda,\theta_0) - \Phi(\lambda)|\right]\right\}/n_0\right\}, \quad (5)$$

where $i$ denotes the $i$th metasurface, $n_0$ is the total number of the meta-atoms of the doublet, and the radius of this doublet is 25.05 μm. Then, the optimal $C(r,\lambda)$ and $F_0$ are found to get the global minimum of 0.8 rad (FOM$_1$) and 0.86 rad (FOM$_2$) by a global optimization algorithm. The theoretical and optimized $\varphi_1$ and $\varphi_2$ at the three designed wavelengths are respectively shown in Figs. 2.

The phase errors of the two metasurfaces are used to further analyze the optimized results, which are defined as:

$$\Delta p_i = \left\{ \sum_{\theta_0=0}^{2\pi} \sum_{r=0}^{R} |\varphi_i(r,\lambda,\theta_0) - \Phi(\lambda)| \right\} / n_{total}, \tag{6}$$

where $i$ denotes the $i$th metasurface, The calculated $\Delta p_1$ are 0.12 rad, 0.29 rad, and 0.39 rad at the wavelengths of 440 nm, 540 nm and 640 nm, respectively. And $\Delta p_2$ are respectively 0.14 rad, 0.31 rad, and 0.41 rad at the three wavelengths. The increased phase errors of the longer wavelengths are due to insufficient phase coverage.

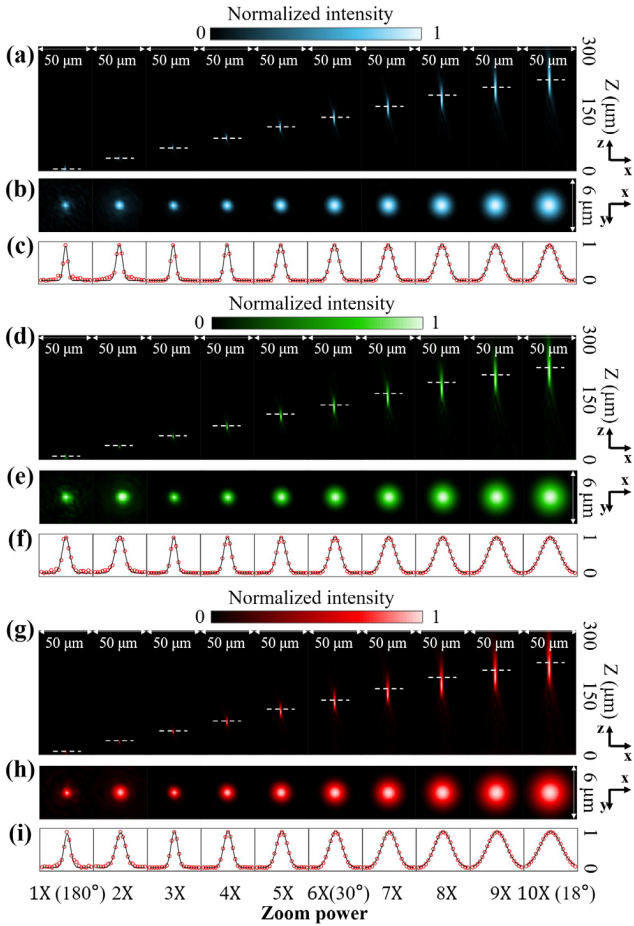

Fig. 3. The achromatic focusing behavior of the doublet. (a-c) The calculated results at the wavelength of 440 nm. (a) The normalized intensity distributions at the $xz$ plane. The actual focal planes are marked with white dash lines. (b-c) The normalized intensity distributions of the central areas of the focal plane and along the $x$-axis. (d-f), (g-i) The calculated results at the wavelengths of 540 nm and 640 nm. The graphs from the first column to the last column correspond to zoom power from 1× to 10×, respectively

Once the two optimized metasurfaces are achieved, we can form the doublet by placing M1 and M2 face to face with an axial distance of 1400 nm. This distance is selected to avoid unwanted dispersion. The Rayleigh Sommerfeld diffraction algorithm is used to calculate the focusing fields based on the near field of the doublet [20].

**Results.** Fig. 3 shows the achromatic focusing effects of the zoom metalens doublet. The normalized intensity distributions at the $xz$ plane, the focal plane, and along the x-axis at the wavelength 440 nm are respectively depicted in Figs. 3(a-c) under the ten rotation angles. The focal lengths changes from 20.47 μm to 218.33 μm, corresponding to the zoom powers from 1× to 10×. As expected, the sizes and depths of the focal spot increase with the increasing zoom power. Figs. 3(d-f), and 3(g-i) respectively show that there are similar results at the wavelengths of 540 nm and 640nm. The overall results reveal the tunability and achromatism of the doublet.

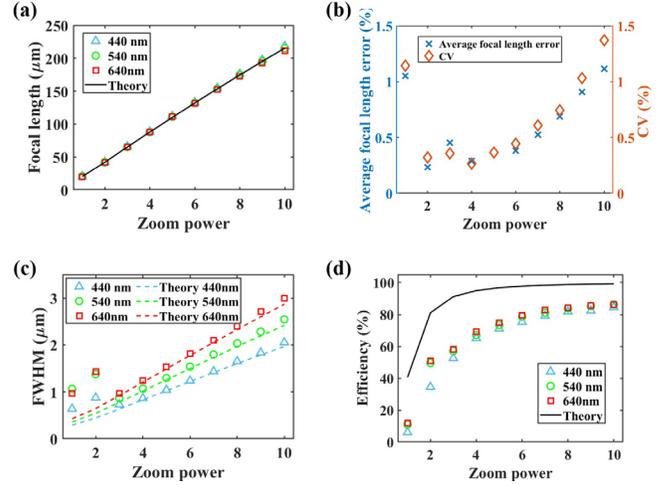

Fig. 4. The quantitative characterizations of the doublet. (a) The focal length versus zoom power. The colored markers and the black line respectively denote denotes the calculated and theoretical values (b) The average focal length error and CV versus zoom power. (c) FWHM versus zoom power. The colored dash lines denote theoretical values. (d) Efficiency versus zoom power at the three wavelengths.

To quantitatively analyze the achromatic focusing performance of the doublet, we take the following indices as the figure of metric, namely the full width at half maxima (FWHM), average focal length error, coefficient of variation (CV=SD/MN×100%) [21], efficiency. Here, SD and MN are the standard deviation and average of the focal length, the average focal length error is defined as the ratio of the average focal length shift to the theoretical value under a certain rotation angle, and efficiency is defined as the ratio of the power within a circular area with a diameter of 3 times the FWHM to the total power at the focal plane.

Fig. 4(a) shows that, under every zoom power, the calculated focal lengths fit well with the theoretical values at the designed wavelengths. These results reveal the achromatic tunability of the doublet. Considering the focal length deviation of the ideal spherical lens (as shown in Supplement), we take the average focal length as the ideal value under a certain zoom power (as shown in Supplement Table S1). To further characterize the achromatic focusing behavior, the concepts of the CV (marked with orange prismatic) and the average focal length errors (marked with sky-blue crosses) are introduced. As illustrated in Fig. 4(b), the average focal length errors and CVs first decrease and then increase with the increasing zoom power. Here, the relatively larger average focal length errors (under zoom powers 1×, 2×, 9×, and 10×) are caused by the spherical phase (as shown in Supplement Fig. S3). All the calculated average focal length errors and CVs are respectively below 1.1 % and 1.3 %. These results further prove the excellent achromatism of the doublet.

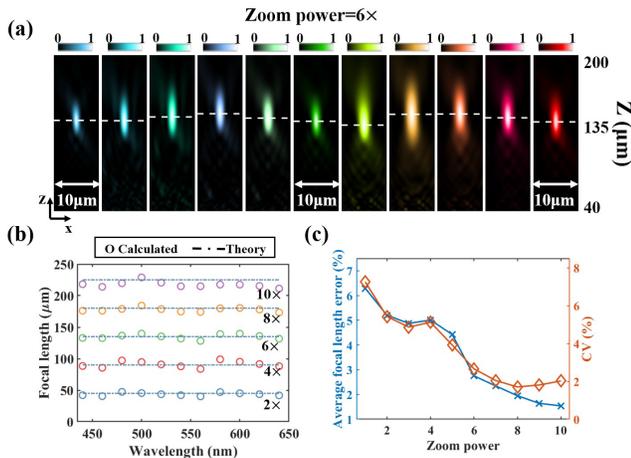

**Fig. 5.** The achromatic focusing behavior of the doublet. in the whole band of 440-640 nm. (a) The normalized intensity distributions at the *xz* plane when the zoom power is 6×. The white dash line represents the actual focal plane. (b) The calculated focal length versus the wavelength when the zoom powers vary from 2× to 10×. (c) The average focal length error and CV versus zoom power.

FWHM is used to evaluate the performance of the focused spots. As is shown in Fig. 4 (c), the FWHMs fit well with the diffraction limits under the zoom powers from 3× to 10×. The slight deviations from the theoretical value under the 1× and 2× zoom powers are caused by the discrepancy between the effective numerical apertures (NA) and the ideal NAs of the doublet. Fig. 4(d) shows that efficiency increases with the increasing zoom power and can reach the maximum value of 86.2% under the 10× zoom power. This behavior can be explained as discussed earlier. Besides, all the calculated efficiencies are lower than the theoretical value due to the imperfect focusing of the spherical lens (as shown in Supplement Fig. S2). Illustrated in Supplement Fig. S4, the non-unity transmittance of M1 and M2 contributes to the lower efficiency at the short wavelength.

We further study the focusing properties at other off-design wavelengths. Fig. 5 (a) describes the focusing performances at the wavelengths from 440 nm to 640 nm in a step of 20 nm when the zoom power is 6×. The focal lengths decrease with the increasing wavelengths from 440 nm to 460 nm due to the dispersion of the doublet [18]. There are similar results in the bands of 480 – 580 nm and 580 – 640 nm. And the maximum focal length shift is 7.77 μm, which is within the minimum depth of focus ( 9.11 μm @440 nm ) in the band of 440– 640 nm. Therefore, there should exist a sharing focal plane where the doublet can achromatically work in this band. Fig. 5 (b) shows that similar achromatism can be found for other zoom powers, and the detailed intensity profiles can be seen in Supplement Fig. S5. As shown in Fig. 5 (c), the calculated average focal length errors and CVs are respectively below 6.28% and 7.25% in this band. These results demonstrate the achromatism and tunability of the doublet in the whole band of 440– 640 nm.

In summary, we have designed an achromatic zoom metalens doublet in the visible band. This doublet can realize a zoom power as large as 10×, featuring an average focal length error below 6.28%, CV below 7.25%, and high efficiency up to 86.2%.

**Funding.** This work is supported by the Natural Science Foundation of China (No.62075073, 62135004 and 62075129), the Fundamental Research Funds for the Central Universities (No. 2019kfyXKJC038), State Key Laboratory of Advanced Optical Communication Systems and Networks, Shanghai Jiao Tong University (No. 2021GZKF007), and Key R & D project of Hubei Province (No. 2021BAA003).

**Disclosures.** The authors declare no conflicts of interest.

**Data availability.** Data underlying the results presented in this paper are not publicly available at this time but may be obtained from the authors upon reasonable request

†These authors contributed equally to this work.
See Supplement 1 for supporting content